\documentclass[a4paper,11pt]{article}
\usepackage{amssymb}
\usepackage{amsmath}
\usepackage{graphicx}
\usepackage{amsbsy}
\usepackage{xr}
\usepackage{bm}
\usepackage{color}
\usepackage{gensymb}
\usepackage{caption}
\usepackage{subcaption}
\usepackage[space]{grffile}
\definecolor{cream}{RGB}{222,217,201}
\usepackage{siunitx}
\DeclareSIUnit[number-unit-product = {\,}]
\cal{cal}
\DeclareSIUnit\kcal{\kilo\cal}
\DeclareSIUnit\kcal{\kilo\joule\per\mole}
\DeclareSIUnit\molar{\mole\per\cubic\deci\metre}
\DeclareSIUnit\Molar{\textsc{m}}
\usepackage{multirow}

\usepackage{setspace}

\usepackage[utf8]{inputenc}
\usepackage[T1]{fontenc}
\usepackage{authblk}

\usepackage[margin=0.7in]{geometry}

\begin{document}

\title{\bf Spontaneous dimensional reduction and novel ground state degeneracy in a simple chain model}

\author[1,2]{Tatjana \v{S}krbi\'{c}\thanks{tskrbic@uoregon.edu}}
\author[3]{Trinh Xuan Hoang\thanks{hoang@iop.vast.ac.vn}}
\author[2]{Achille Giacometti\thanks{achille.giacometti@unive.it}}
\author[4]{Amos Maritan\thanks{amos.maritan@unipd.it}}
\author[1]{Jayanth R. Banavar\thanks{{\it corresponding author:} banavar@uoregon.edu}}

\affil[1]{\small \it Department of Physics and Institute for Fundamental Science, University of Oregon, Eugene, OR 97403, USA}

\affil[2]{\small \it 
Dipartimento di Scienze Molecolari e Nanosistemi,
Universit\`{a} Ca' Foscari di Venezia
Campus Scientifico, Edificio Alfa,
via Torino 155, 30170 Venezia Mestre, Italy}

\affil[3]{\small \it Center for Computational Physics Institute of Physics,
Vietnam Academy of Science and Technology,
10 Dao Tan St., Hanoi, Vietnam}

\affil[4]{\small \it Dipartimento di Fisica e Astronomia,
Universit\`{a} di Padova and INFN
via Marzolo 8, 35131 Padova, Italy}

\date{}


\maketitle

\begin{abstract}Chain molecules play a key role in the polymer field and in living cells. 
Our focus is on a new homopolymer model of a linear chain molecule subject to 
an attractive self-interaction promoting compactness. We analyze the model using 
simple analytic arguments complemented by extensive computer simulations. We find 
several striking results: there is a first order transition from a high temperature 
random coil phase to a highly unusual low temperature phase; the modular ground states 
exhibit significant degeneracy; the ground state structures exhibit spontaneous dimensional reduction 
and have a two-layer structure; and the ground states are assembled from secondary motifs 
of helices and strands connected by tight loops. We discuss the similarities and notable differences 
between the ground state structures (we call these PoSSuM -- Planar Structures with Secondary Motifs) 
in the novel phase and protein native state structures. 
\end{abstract}



\newpage

Materials made up of polymer chains are ubiquitous in everyday life and in industry. Here we study a simple model 
of a chain with tuned interactions, which yields very unusual behavior of the ground state conformation(s) 
of the chain. Intriguingly, even though the chain lives in three dimensional space, it sacrifices exploring all three dimensions and spontaneously becomes a two layer structure in order to benefit from the maximal number of contacts. The system is therefore Euclidean two-dimensional in its ground state and should not be confused with the fractal dimension of two adopted by a random walk. Furthermore, this layered structure exhibits strands and helices that are able to interchange with each other resulting in a huge ground state degeneracy. Quite remarkably, this behavior is robust on changing all but one parameters of the model (this crucial parameter needs to be held fixed to maintain the interchangeability of helices and strands). Finally, our results are {\it not} artifacts of finite size effects. 

We consider a chain \cite{Rubinstein_Colby,Doi_Edwards} of $N$ spheres with diameter $\sigma$ 
tethered to each other in a railway train topology with a fixed bond length $b$. In what follows, 
all length scales are measured in units of the bond length $b$, which we set equal to $1$ without loss 
of generality. A chain is inherently anisotropic because of tethering while an individual sphere is 
isotropic. In order to avert this spurious symmetry, we consider two modifications. 
First, adjacent spheres overlap \cite{clementi,magee_helix,coluzza} -- $\sigma$ is allowed to be larger than $1$. 
Second, side spheres \cite{klimov_thirumalai,muthukumar_pnas,hoang_PNAS_2009,soft_matter_2016,Skrbic_JCP_2016,elixir,local_symmetry} 
of diameter ${\sigma}_s$ are attached (tangent to the main chain sphere in the negative normal direction in the local Frenet frame) to each of the main chain spheres except the first and the last. 
Other than the adjoining main chain spheres along the chain, none of 
the main chain and side spheres is allowed to overlap. We impose a pairwise attractive interaction 
between the main chain spheres within the range of attraction $R$, and separated by at least $4$ along the sequence, with an energy scale of $\epsilon$ (set equal to 1). We ascribe a distinct weaker energy 
of $x=-1/2$ for a pair of main chain spheres separated by exactly $3$ spheres along the sequence and within the 
range of interaction $R$. This tuning yields the very special highly degenerate ground states assembled from building blocks of helices and strands. There is no attractive interaction for main chain spheres with sequence separation less than $3$. All energies 
(including the temperature) are measured in units of the depth of the attractive square well 
potential $\epsilon$. 

\begin{figure}[htpb]
\centering
\captionsetup{justification=raggedright,width=\linewidth}
\includegraphics[width=\linewidth]{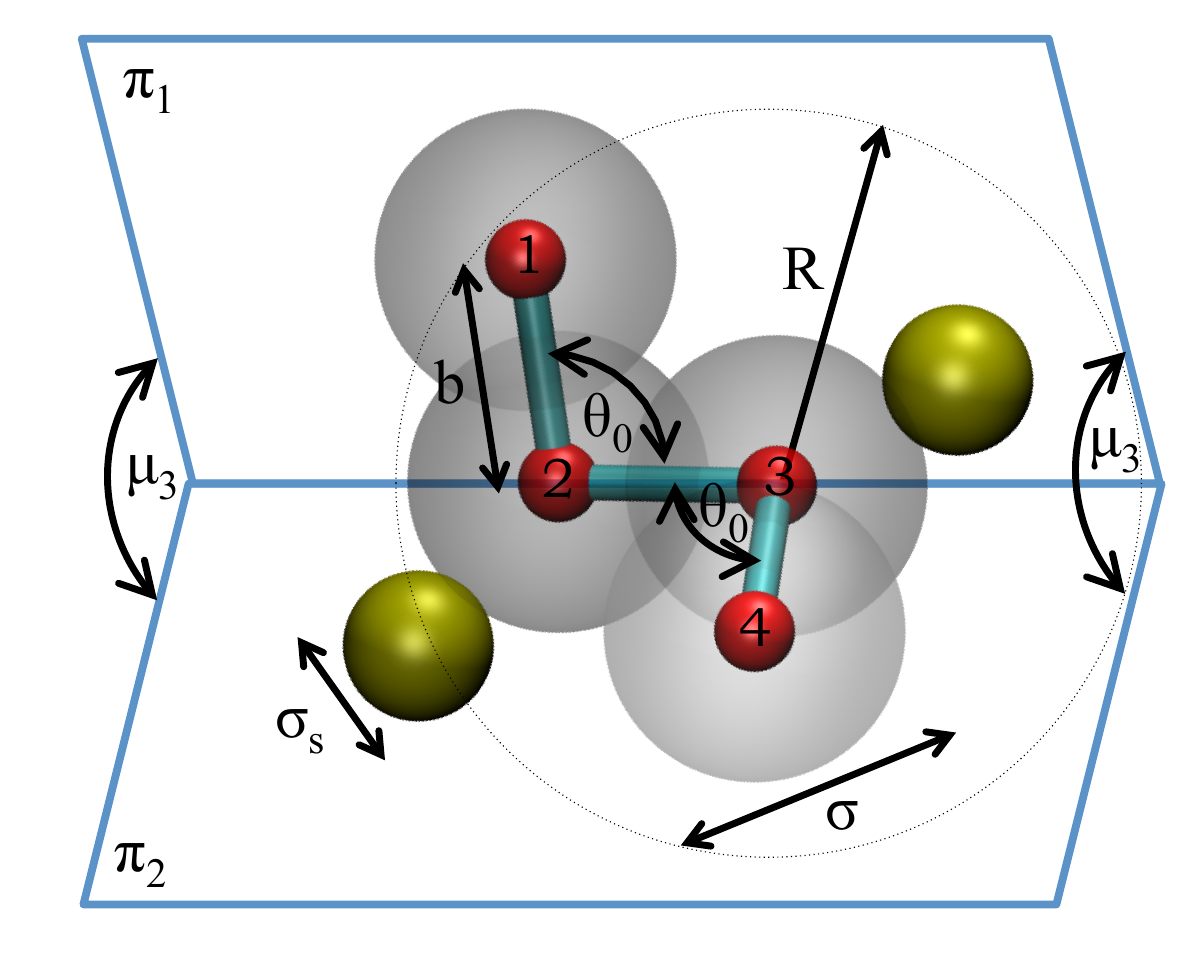}
\caption{Coordinate system used in our study. Here we show $4$ consecutive main chain spheres along the chain. 
The bond length $b=1$ sets the length scale of the model. The main chain spheres (shown in grey with centers marked 
by little red spheres) have a diameter $\sigma$ larger than $b$ leading to an overlap of successive spheres. 
The yellow side spheres of diameter ${\sigma_s}$ are attached to the main chain spheres tangentially in the negative 
normal direction. The angles between successive bonds are held constant at a value ${\theta}_0$. The range of attractive 
interactions is denoted by $R$. ${\mu}_3$ is the dihedral angle between the planes (${\pi}_1$ and ${\pi}_2$) defined by sphere 
centers ($1$,$2$,$3$) and ($2$,$3$,$4$), respectively. It is also the angle between successive binormal vectors in a Frenet coordinate 
system at sphere centers $2$ and $3$. 
\label{fig:possum_model}}
\end{figure}

The coordinate system we employ is shown in Figure \ref{fig:possum_model}. For fixed bond length, a chain conformation is specified by two angles $\theta$ and $\mu$. $\theta$ is a measure of bond bending, a straight conformation has $\theta =\pi$. Two distinct kinds of bending energy penalties have been commonly employed in the literature: an energy cost proportional 
to ${\cos}^2(\theta /2)$; or zero cost when $\theta > {\theta}_0$ and infinite cost otherwise 
\cite{seaton_semiflexible, chain_stiffness}. 
Here we use a third approach and make the simplification of fixing $\theta=\theta_0$ resulting in $\theta$ no longer being a variable of the model but just a parameter. This can be thought of as 
$\theta$ being pinned at its minimum allowed value due to the tendency for compactness and yields a freely
rotating chain \cite{Rubinstein_Colby,Doi_Edwards}. The angle between successive 
binormals, $\mu$, is the second angular coordinate and is the dihedral or torsional angle. For simplicity, we do not incorporate a torsional rigidity energy, 
conventionally chosen to be proportional to $\sin^2(\mu/2)$ \cite{marenduzzo_maritan}, or any other explicit energy 
dependence on $\mu$. The parameters of the model are $\sigma$, ${\sigma}_s$, $R$, $x$, and $\theta_0$. The PoSSuM phase 
is centered around $\sigma=4/3$, ${\sigma}_s=2/3$, $R=8/5$, $x=-1/2$, and ${\theta}_0 = 97^{\circ}$. The phase is robust 
to small variations (of the order of 10\%) in these parameter values with one exception. As we shall see, the rich 
non-trivial degeneracy of ground state conformations requires $x$ to be $-1/2$.

\begin{figure}[htpb]
\centering
\captionsetup{justification=raggedright,width=\linewidth}
\includegraphics[width=\linewidth]{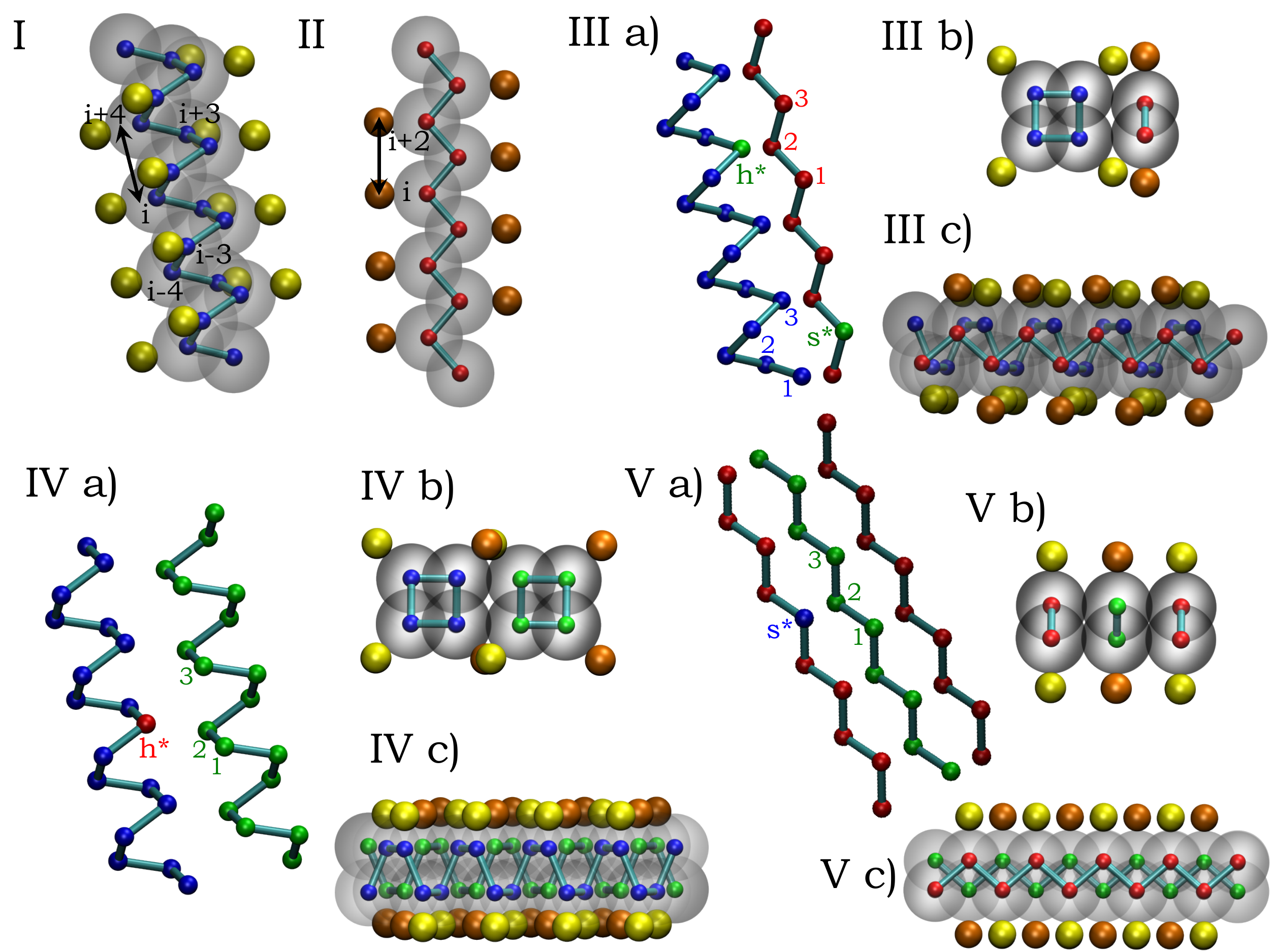}
\caption{Sketches of ideal assembly of secondary motifs. Panels I and II show snapshots of an individual helix and strand with
perfect commensurability attributes. The model has $\theta = {\theta}^{*} = 98.213^{\circ}$ and the optimal values of 
$\mu = \mu^{*} = 41.410^{\circ}$ and $180^{\circ}$ for the helix and strand respectively. There is commensurability both in the 
number of main chain spheres per repeat unit ($4$ for the helix and $2$ for the strand) as well as equality of the lengths of the 
repeat units (helix ($i$,$i+4$) distance = strand ($i$,$i+2$) distance = $1.512$ in units of the bond length). The small spheres 
outline the backbone of the chain, the large grey spheres are the main chain spheres (diameter = $4/3$), the little yellow (for helix) 
(orange, for strand) spheres are the side spheres (diameter = $2/3$) pointed in
the negative normal direction. A solitary strand has no local contacts, whereas the local energy score per interior helix main chain sphere
is $-3$ arising from energies of $-1$,$-1/2$,$-1/2$, and $-1$ for $4$ contacts of sphere $i$ with spheres $i-4$,$i-3$,$i+3$, $i+4$, 
respectively. Panels III, IV, and V show the perfect fit of a helix with a strand; two helices of opposite chirality; and three strands. 
The three sub-panels show three views (front, top, and side) of each of these assemblies, in particular, revealing the two layer structure. 
Panels III (a), IV (a), and V (a) show a trace of the backbone of the secondary motifs (without main chain or side spheres) and the three 
non-local contacts made by each internal sphere of the secondary motif. For a strand, the total number of non-local contacts is $6$ 
($3$ from one side and $3$ from the other), whereas for a sphere in a helix, there are four local contacts (yielding a favorable energy 
of $-3$ -- recall $x=-1/2$) and three non-local contacts resulting in a net favorable energy of $-6$. Indeed, each of the interior spheres 
of any of these assemblies has a net energy of $-6$ yielding the highly degenerate ground state. Our analysis of the PoSSuM phase in 
this paper is carried out with $\theta = 97^{\circ}$ in the vicinity of ${\theta}^{*}$. The flexibility in the $\theta$ and $\mu$ 
angles allows an excellent match of the ground state PoSSuM structures with the perfect commensurability structures shown here.
\label{fig:ideal_structures}}
\end{figure}

The high temperature phase is a random coil and exhibits the familiar Flory scaling \cite{Rubinstein_Colby,Doi_Edwards} 
of the end-to-end distance and radius of gyration as a function of chain length ($\approx N^{0.6}$). We have verified 
this with simulations for $N$ up to $1024$ (results not shown). At low temperatures, the chain adopts a compact conformation 
to maximize the number of attractive main sphere contacts, while respecting the chain connectivity and steric interactions. 
Akin to a periodic crystal, the simplest structured conformations of a chain arise from a repeat of the $\mu$ values along 
the chain. For the state point above, any uniform $\mu \geq \mu_{min}=36.45^{\circ}$ yields clash-free chains of indefinite 
length. A repeat structure with $\mu=\mu_{min}$ results in the most tightly wound helix with the maximum number of local 
contacts -- a sphere in the interior of the helix $i$ has $4$ contacts with spheres separated in sequence by $-4$, $-3$, $+3$, 
and $+4$, yielding a total attractive energy of $-3$ units per interior sphere
(Figure \ref{fig:ideal_structures}, Panel I). The same $4$ contacts remain until the $\mu$ 
value of $45.285^{\circ}$. One obtains a two-dimensional strand, when $\mu$ is close 
to $\pi$ (Figure \ref{fig:ideal_structures}, Panel II). The strand does not have any within-strand contacts and has zero attractive 
energy per interior sphere. We note the key observation, made in the protein context \cite{structural_hybrids,rose_GNR}, that sterics 
inhibits structural hybrids and promotes structures with repeat $\mu$ values. 
Indeed, we find in simulations that a random equal choice of helix-strand $\mu$-values in a chain of length 40 yields 
fewer than $7\%$ of viable chains with no steric clashes.

Armed with the insight that individual helices and strands are building blocks of the PoSSuM ground state structures, we 
proceed to work out their harmonious packing. We seek commensurability of the repeat structures -- the pitch of the helix and 
the distance between the ($i$,$i+2$) spheres or corresponding pitch in a strand, which is equal to $2b\sin(\theta/2)$; 
and the number of spheres per turn in a helix and the number of spheres per turn in a strand, equal to $2$. 
One can readily work out the conditions for perfect 
commensurability: $\theta^* = 2\tan^{-1}(\sqrt{4/3})\sim 98.213^\circ$ for our model with $\mu^* = 41.410^\circ$ for the 
helix and $180^\circ$ for the strand. For these choices, the special helix has exactly $4$ main chain spheres per turn 
(with perfect commensurability with the strand) and its pitch (the $i$,$i+4$ distance) exactly matches the ($i$,$i+2$) distance 
or the corresponding pitch of the planar strand. Figure \ref{fig:ideal_structures} (Panels III--V) illustrates the ideal packing of 
secondary motifs that yield a two layer idealized structure. Helices of opposing chiralities 
tend to pack much better than helices of the same chirality, which cannot make the same number of contacts without creating steric 
clashes between side spheres (not shown).  It is important to note that this ideal commensurate phase point lies within the basin 
of the PoSSuM phase. A single helix has an energy of $-3$ corresponding to $4$ intra-helical contacts (recall our choice of $x=-1/2$) 
and each sphere in the helix has exactly three inter-helical contacts (be it with a single partner helix of opposite chirality 
or a strand) yielding a net energy per sphere of $-6$. A single strand has no attractive energy on its own but accumulates a 
favorable energy of $-6$ per sphere with a contribution of $-3$ from each of its two partners, be it another strand or a helix.

The exact degeneracy of structures in the PoSSuM phase is broken when $x$ deviates from the value of $-1/2$. We can readily work out 
the ground state phases surrounding the PoSSuM phase on varying the model parameters. Too large an attraction range leads to a globular 
phase, whereas the converse yields a sheet phase because the inter-helical contacts are disrupted. Too large a main chain sphere size 
results in steric clashes whereas the converse leads to disruption of the intra-helical contacts, again leading to the sheet phase. 
Too small a side-sphere size yields a globule phase whereas too large a side-sphere size disrupts inter-helical contacts with the 
sheet again emerging as the winning phase. Finally, too small a $\theta$ angle does not allow for a helix with $4$ spheres per turn 
thereby promoting the sheet phase. The sheet phase is also the phase of choice for too large a $\theta$ angle because of the disruption of 
intra-helical contacts -- the distance between the main chain spheres  $i$ and $i+4$  
becomes prohibitively large. In this way, the PoSSuM phase is seen to be nestled between the globular phase and the sheet phase at low 
temperatures, thereby conferring on it the sensitivity associated with being in the vicinity of a phase transition. The robustness of 
the sheet phase is due to the ability to place strands at an optimal distance from each other with no major issues pertaining to 
side-sphere clashes. The PoSSuM phase exists in a Goldilocks window of parameter space characterized by a delicate combination of 
steric constraints both from the main and side spheres, the optimal range of attraction and especially the fine-tuning of 
the attractive interaction, as well as the fixed bond bending angle allowing for commensurability. Interestingly, the PoSSuM 
phase exists only when the main chain spheres overlap thereby removing the spurious spherical symmetry. 

We now turn to a brief description of our computer simulations before presenting the results. The extensive search for ground state 
configurations was performed using Wang-Landau microcanonical Monte Carlo (MC) simulations \cite{wang_landau_prl} with no low energy 
cut-off. The density of states g(E) that are visited along the simulation was iteratively built by filling consecutive energy 
histograms. The acceptance probability was chosen to promote moves toward less populated energy states thus providing for increasing 
flatness of energy histograms with the length of the simulation. The density of states g(E), using which the thermodynamics was 
calculated, was obtained employing cut-off values for energy histograms within 2\% of the pre-determined value of the ground state 
energy of the system. In all cases 28-30 levels of iterations were carried out with a flatness criterion in each iteration of at least 80\%, ensuring convergence of the results \cite{taylor_jcp_2009,taylor_pre_2009,taylor_jcp_2016}. The set of MC moves were modified to maintain $\theta$ constant and included both local-type 
moves, such as 
single-sphere crankshaft, reptation, and end-point moves, as well as non-local type moves, such as the pivot \cite{madras_sokal}.

\begin{figure}[htpb]
\centering
\captionsetup{justification=raggedright,width=\linewidth}
\includegraphics[width=\linewidth]{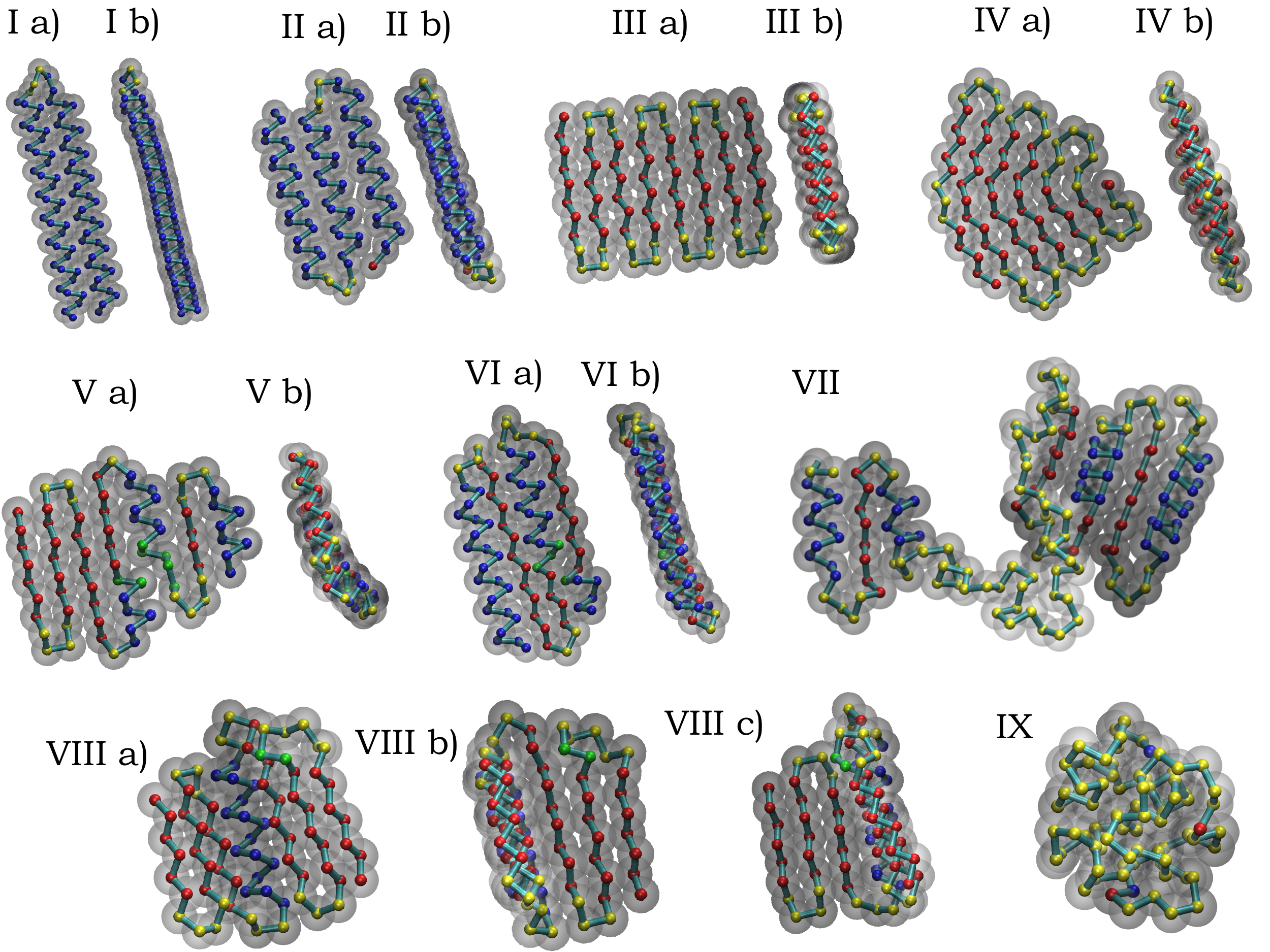}
\caption{Gallery of conformations. All the panels depict conformations of chain length $N=80$ except for VII, which shows domain 
formation for a $N=160$ chain. Panel IX is a globular conformation for the case of a chain with no side spheres. The presence of 
side spheres results in nearly degenerate PoSSuM conformations of two helices (I), three helices (II), a sheet assembled from strands (III), 
a curved sheet (IV), conformations which include a switch between a helix and strand (V and VI), and a pivotal helix that links 
two sheets at right angles to each other (VIII). For several panels, two or three perspectives are shown to illustrate the tight packing 
and the layered structure. The grey circles depict the main chain spheres. The side spheres are not shown. The little circles lie 
at the centers of the main chain spheres. As a guide to the eye, we color a helix sphere center blue, a strand sphere center red, a turn sphere center yellow and a hybrid sphere center in the vicinity of a helix-strand switch green. These assignments are made based on $\mu$ values, the number of main chain spheres per turn, and the number of local and non-local contacts. 
\label{fig:possum_gallery}}
\end{figure}

Figure \ref{fig:possum_gallery} shows a gallery of low energy conformations (for $N=80$) in the PoSSuM phase along with evidence for 
domain formation in simulations of $N=160$. There are deviations from the optimal energy of $-6$ per interior sphere due to turns 
which connect the secondary elements together as well as significant edge effects occurring at the boundaries. A remarkable feature 
of the gallery is the distinct topologies of the degenerate ground state structures. The degeneracy is further enhanced by the possibility 
of a coordinated conversion of a helix to a strand and vice-versa while maintaining the energy. The PoSSuM phase and our model is 
distinct from that used in an earlier study that identified an elixir phase of matter \cite{elixir, local_symmetry}. The previous model 
did not have a fixed $\theta$ angle. However moderate $\theta$ angles were promoted because of an $i-i+2$ attractive interaction. Also, 
$x$ was set equal to $1$ in that model favoring the helix over the sheet. 

\begin{figure}[htpb]
\centering
\captionsetup{justification=raggedright,width=\linewidth}
\includegraphics[width=\linewidth]{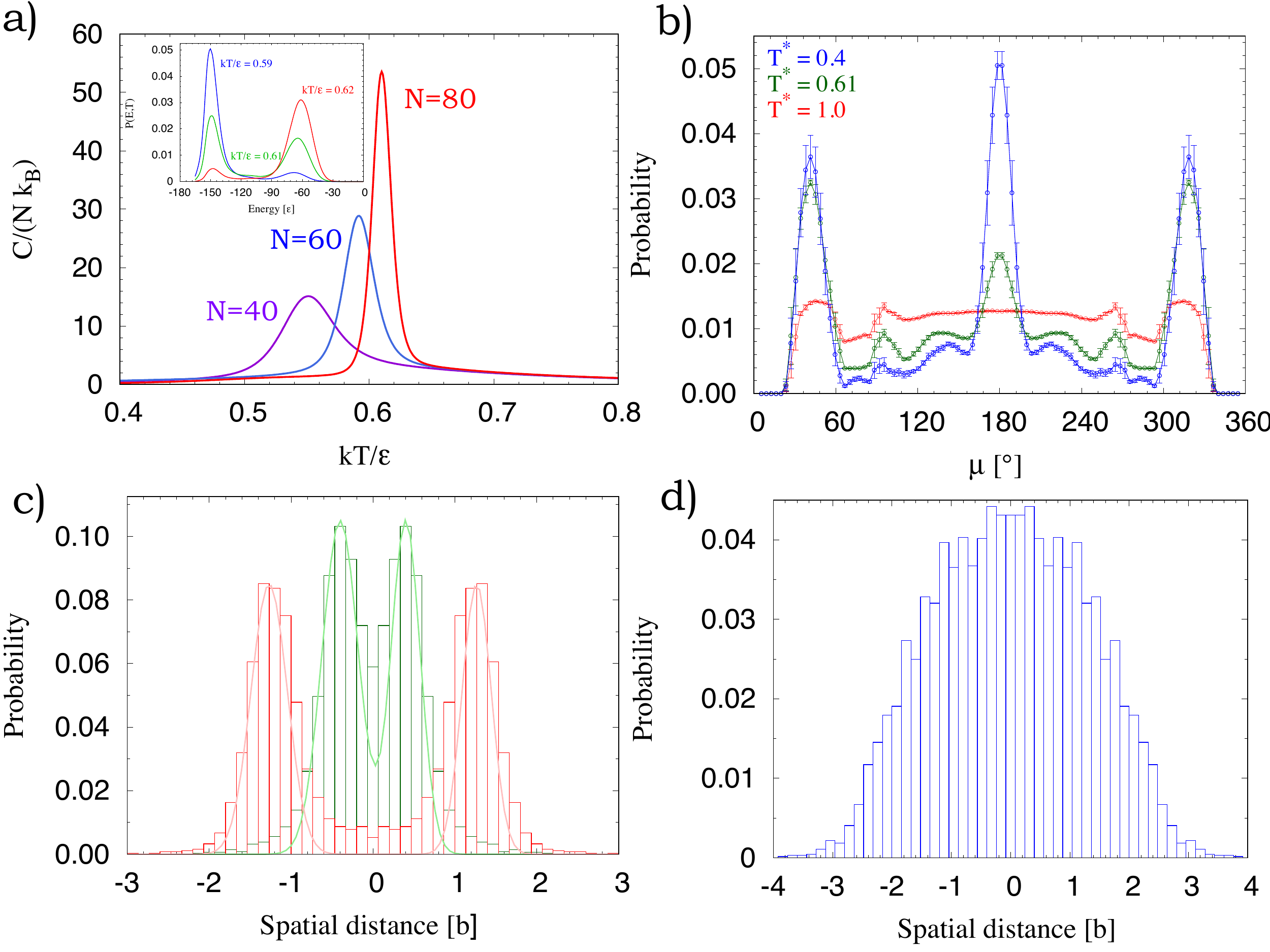}
\caption{Temperature dependence of the PoSSuM model. $a)$ Plot of the specific heat as a function temperature for three chain 
lengths. The peak sharpens and increases approximately linearly with $N$ suggesting that the transition is first order. The inset 
is a plot of the distribution of the canonical energy for $N=80$ for three temperatures in the vicinity of the transition. $b)$ 
The distribution of $\mu$ at three temperatures. At a temperature of $1.0$, there is already a signature of incipient secondary motifs. 
The three peaked structure at the temperature of $0.4$ underscores the presence of secondary motifs. $c)$ The two layered structure 
of the PoSSuM ground states is shown by the green histogram fitted as the sum of two Gaussians. The red histograms indicate the 
spatial spread of the side spheres. In order to determine the direction perpendicular to the layers, we used the eigenvector 
corresponding to the smallest eigenvalue of the moment of inertia matrix. The graph is an average over $40$ distinct ground state PoSSuM 
conformations of $N=80$. $d)$ is the corresponding plot for the globular phase that is obtained when the side chains are removed from the 
model. Note the lack of layering and the higher spread of the main chain spheres.
\label{fig:possum_temperature}}
\end{figure}

Figure \ref{fig:possum_temperature} presents data pertaining to the nature of the phase behavior of the PoSSuM model. Panel (a) is a 
plot of the specific heat versus the temperature for three different chain lengths. The specific heat peak grows approximately 
linearly with $N$ as expected for a first order phase transition \cite{taylor_pre_2009,doniach_orland,los_rios}. 
The inset of the figure shows the canonical energy $P(E,T)$, which 
exhibits the characteristic bimodal shape at a first order transition. Panel (b) shows a histogram of $\mu$ values for three different 
temperatures, one in the high temperature phase, another in the vicinity 
of the transition and the third in the low temperature phase. At low temperatures, the peaks occur at the $\mu$-values of the helix, 
the strand, and the turns. There is a signal of the formation of secondary motifs in the histogram of $\mu$ angles even at 
a temperature more than $50\%$ larger than the transition temperature. Panel (c) shows the layered structure of 
the PoSSuM ground states. The absence of sharp layering is readily accounted for because the layer separation depends on the 
secondary motifs adjacent to each other. The side spheres lie on the opposite ends of the layers and prevent attractive interactions 
with a putative third layer and growth in the third dimension. Finally, Panel (d) is the spatial configuration associated with a 
ground state gallery of the globular phase obtained by taking the PoSSuM model and eliminating the side-spheres. 

\begin{figure}[htpb]
\centering
\captionsetup{justification=raggedright,width=\linewidth}
\includegraphics[width=\linewidth]{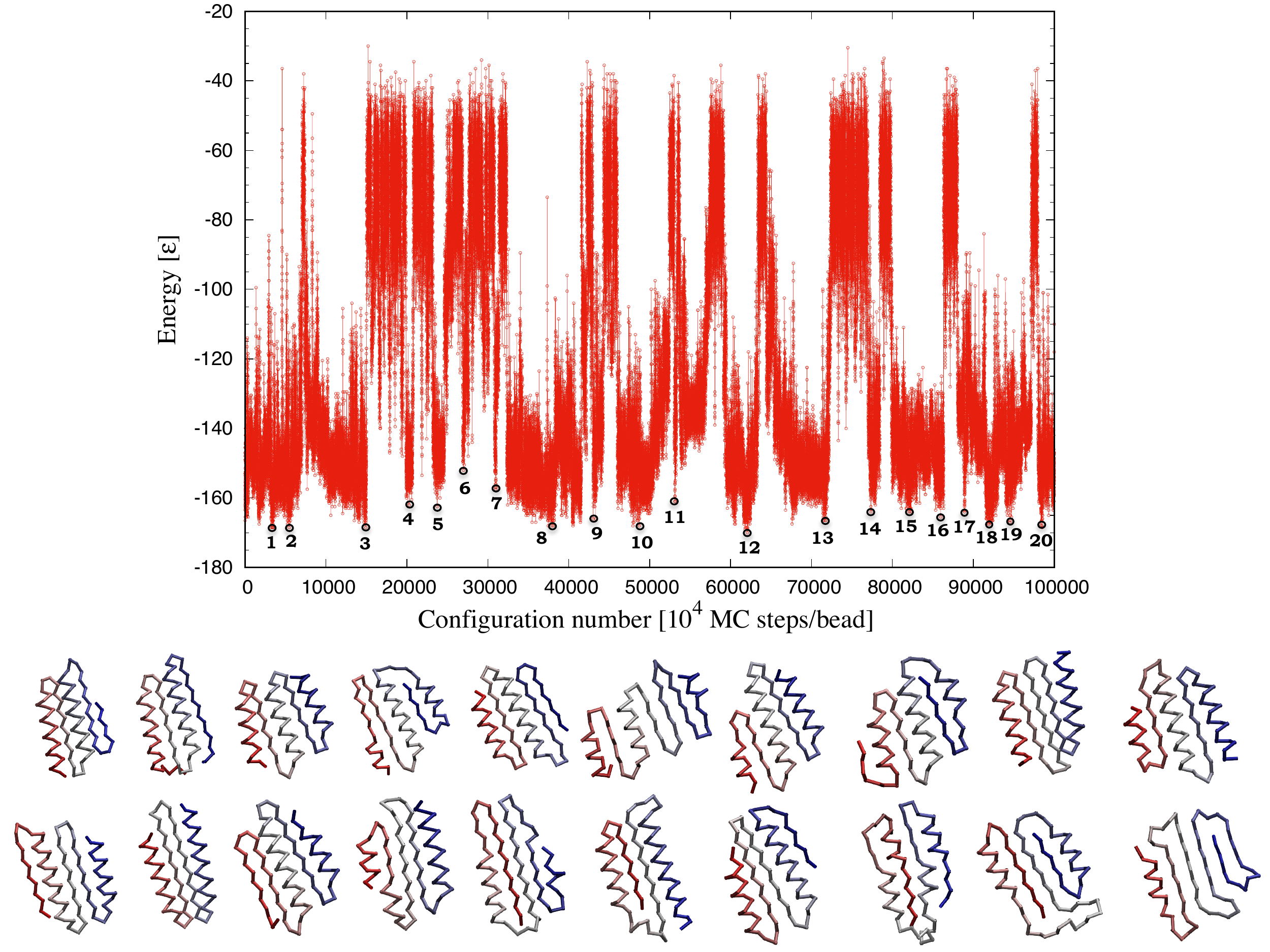}
\caption{A single constant temperature Monte Carlo run for $N=80$ at a reduced temperature of 0.595, around 2\% below the transition 
temperature. The energy is plotted as a function of Monte-Carlo time and shows many switches between the unfolded state and the 
PoSSuM state. Twenty well-formed nearly degenerate PoSSuM configurations with distinct topologies are shown at the bottom of the figure in the same order as their appearance in the Monte-Carlo run.
\label{fig:possum_switching}}
\end{figure}

Figure \ref{fig:possum_switching} shows the energy as a function of MC time in a single long run at a temperature equal to approximately $98\%$ of 
the transition temperature. It shows multiple switching between the high temperature and low temperature phases as expected at a 
first order transition. Furthermore the low energy structures (we show 20 of these visited in just the single run) are very well-formed 
modular structures made up of helices and strands and having energies within $10\%$ of the lowest energy conformation obtained in  
detailed Wang-Landau computer simulations \cite{wang_landau_prl}. 
Finally, the PoSSuM phase is not a finite size effect. In the thermodynamic limit, it is straightforward to deduce that the geometries 
of the degenerate ground states are still essentially planar with a thickness of two layers with both the length and width scaling as 
$\sim \sqrt{N}$.

Even though our model has features reminiscent of native structures of proteins
\cite{creighton,bahar_jernigan_dill,lesk,comment} 
(the modular building blocks of helices and strands arising from the presence of side-chains is one striking commonality), there are 
essential differences. Proteins are made up of twenty types of naturally occurring amino acids. Here instead we consider a homopolymer 
model. The ground state is nevertheless found to be highly degenerate. Thus, upon adding sequence heterogeneity, a given sequence has a large predetermined menu of structures to choose its ground state from. 
Proteins do not have a fixed $\theta$ angle unlike in our simplified model. The number of amino acids per turn in a protein is 
approximately $3.6$. Here we have a nice integer of $4$. There is chiral symmetry breaking in a protein unlike in our model. In fact, the 
assembly of right-handed and left-handed helices in the PoSSuM phase cannot happen in protein structures. There is also an important 
difference in the assembly of strands into sheets. The need for close packing promotes the assembly of out of phase strands in the 
PoSSuM phase, whereas strands making up a protein tend to be in phase. Protein structures are three dimensional and proteins misfold 
and aggregate into amyloid. The PoSSuM phase, in contrast, is beautifully packed in two layers. 

Francis Crick noted \cite{crick}: {\it Physicists are all too apt to look for the wrong sorts of generalizations, to concoct 
theoretical models that are too neat, too powerful, and too clean. Not surprisingly, these seldom fit well with data.} While our model is, 
in fact, a too neat and clean model, it does not and nor is it meant to describe proteins, the amazing molecular machines of life. 
It nevertheless carries important lessons for physics. What we have demonstrated here is the existence of a novel phase of matter 
in the context of a simplified chain model. The continuous $\theta$-transition in standard chain models is replaced by a first order 
transition. The low temperature PoSSuM phase is characterized by a spontaneous dimensional reduction with the ground states occupying 
two layers. There are numerous non-trivially related ground states arising from the modular building blocks of helices and strands. 
One cannot help but wonder what useful hints the PoSSuM phase might offer for understanding the magnificent protein native state structures.\\

{\bf Acknowledgements:} We are indebted to George Rose for collaboration and inspiration. We are very grateful to Pete von Hippel for his warm hospitality and to him and Brian Matthews for stimulating conversations.\\ 

{\bf Funding:} This project received funding from the European Union’s Horizon 2020 research and innovation program under the Marie Sk\l odowska-Curie Grant Agreement No 894784. The contents reflect only the authors’ view and not the views of the European Commission. Support from the University of Oregon (through a Knight Chair to JRB), Vietnam National Foundation for Science and Technology Development (NAFOSTED) under grant number 103.01-2019.363 (TXH), University of Padova through ``Excellence Project 2018'' of the Cariparo foundation (AM), MIUR PRIN-COFIN2017 Soft Adaptive Networks grant 2017Z55KCW and COST action CA17139 (AG) is gratefully acknowledged. The computer calculations were performed on the Talapas cluster at the University of Oregon. {\bf Conflict of interest}: The authors declare that they have no conflict of interest.\\

{\bf Author Contributions:} T\v{S} and JRB conceived the ideas for the calculations. T\v{S} carried out the calculations. JRB wrote the paper. All authors participated in understanding the results 
and reviewing the manuscript.\\


\begin{thebibliography}{999}

\bibitem{Rubinstein_Colby}
M.~Rubinstein and R.~H. Colby, {\em Polymer Physics (Chemistry)}.
\newblock Oxford University Press, 1~ed., 2003.

\bibitem{Doi_Edwards}
M.~Doi and S.~F. Edwards, {\em The theory of polymer dynamics}.
\newblock Clarendon Press, New York, 1993.

\bibitem{clementi}
C.~Clementi, A.~Maritan and J.~Banavar,
``Folding, design, and determination of interaction potentials using off-lattice
dynamics of model heteropolymers'',
{\em Phys. Rev. Lett.} {\bf 81}, 3287--3290 (1998).

\bibitem{magee_helix}
J.~E.~Magee, V.~R.~Vasquez and L.~Lue,
``Helical structures from an isotropic homopolymer model'',
{\em Phys. Rev. Lett.} {\bf 96}, 207802 (2006).

\bibitem{coluzza}
I.~Coluzza, ``A Coarse-Grained Approach to Protein Design:
Learning from Design to Understand Folding'',
{\em PLoS One} {\bf 6}, e20853 (2011).

\bibitem{klimov_thirumalai}
D.~K.~Klimov and D.~Thirumalai,
``Cooperativity in protein folding: from lattice models with
side-chains to real proteins'',
{\em Folding \& Design} {\bf 3}, 127--139, (1998).

\bibitem{muthukumar_pnas}
M.~Muthukumar and C.~Y.~Kong,
``Simulation of polymer translocation through protein channels'',
{\em Proc. Natl. Acad. Sci. USA} {\bf 103}, 5273--5278 (2005).

\bibitem{hoang_PNAS_2009}
J.~R. Banavar, M.~Cieplak, T.~X. Hoang and A.~Maritan,
``First-principles design of nanomachines'', {\em Proc. Natl. Acad. Sci. USA} {\bf 106},
6900--6903 (2009).

\bibitem{soft_matter_2016}
T.~\v{S}krbi\'{c}, A.~Badasyan, T.~X. Hoang, R.~Podgornik and A.~Giacometti,
``From polymers to proteins: the effect of side chains and broken symmetry
on the formation of secondary structures within a Wang–Landau approach'',
{\em Soft Matter} {\bf 12}, 4783--4793 (2016).

\bibitem{Skrbic_JCP_2016}
T.~\v{S}krbi\'{c}, T.~X.~Hoang and A.~Giacometti,
``Effective stiffness and formation of secondary structures in a protein-like model''.
{\em J. Chem. Phys.} {\bf 145}, 084904 (2016).

\bibitem{elixir}
T.~{\v{S}}krbi{\'c}, T.~X. Hoang, A.~Maritan, J.~R. Banavar and A.~Giacometti,
``The elixir phase of chain molecules'', {\em Proteins: Structure, Function,
and Bioinformatics} {\bf 87}, 176--184 (2019).

\bibitem{local_symmetry}
T.~\v{S}krbi\'{c}, T.~X. Hoang, A.~Maritan, J.~R. Banavar and A.~Giacometti,
``Local symmetry determines the phases of linear chains: a simple model for
the self-assembly of peptides'', {\em Soft Matter} {\bf 15}, 5596--5613 (2019).

\bibitem{seaton_semiflexible}
D.~T.~Seaton, S.~Schnabel, D.~P.~Landau and M.~Bachmann,
``From flexible to stiff: Systematic analysis of structural phases for single
semiflexible polymers'',
{\em Phys. Rev. Lett.} {\bf 110}, 028103 (2013).

\bibitem{chain_stiffness}
T.~\v{S}krbi\'{c}, J.~R. Banavar and A.~Giacometti,
``Chain stiffness bridges conventional polymer and bio-molecular phases'',
{\em J. Chem. Phys.} {\bf 151}, 174901 (2019). and references therein

\bibitem{marenduzzo_maritan}
D.~Marenduzzo, C.~Micheletti, H.~Seyed-allaei, A.~Trovato and A.~Maritan,
``Continuum model for polymers with finite thickness'',
{\em J. Phys. A: Math. Gen.} {\bf 38}, L277--L283 (2005).

\bibitem{structural_hybrids}
N.~C.~Fitzkee and G.~D.~Rose, ``Steric restrictions in protein folding:
an alpha-helix cannot be followed by a contiguous beta-strand'',
{\em Prot. Sci.} {\bf 13}, 633--639 (2004).

\bibitem{rose_GNR}
G.~D.~Rose, ``In Memoriam: Professor G.N. Ramachandran (1922 -- 2001)'',
Perspective -- {\em Proteins: Structure, Function, and Bioinformatics} {\bf 10},
1691--1693 (2001).

\bibitem{wang_landau_prl}
F.~Wang and D.~P.~Landau, ``Efficient, Multiple-Range Random Walk Algorithm
to Calculate the Density of States'',
{\em Phys. Rev. Lett.} {\bf 86}, 2050 (2001).

\bibitem{taylor_jcp_2009}
M.~P.~Taylor, W.~Paul and K.~Binder,
``Phase transitions of a single polymer chain: A Wang-Landau simulation study'',
{\em J. Chem. Phys.} {\bf 131}, 114907 (2009).

\bibitem{taylor_pre_2009}
M.~P.~Taylor, W.~Paul and K.~Binder,
``All-or-none proteinlike folding transition of a flexible homopolymer chain'',
{\em Phys. Rev. E} {\bf 79}, 050801 (2009).

\bibitem{taylor_jcp_2016}
M.~P.~Taylor, W.~Paul and K.~Binder,
``On the polymer physics origins of protein folding thermodynamics'',
{\em J. Chem. Phys.} {\bf 145}, 174903 (2016).

\bibitem{madras_sokal}
N.~Madras and A.~D.~Sokal,
``The pivot algorithm:  A highly efficient Monte Carlo method for the self-avoiding walk'',
{\em J. Stat. Phys.} {\bf 50}, 109--186 (1988).

\bibitem{doniach_orland}
S.~Doniach, T.~Garel and H.~Orland, ``Phase diagram of a semiflexible polymer
chain in a theta solvent: Application to protein folding'',
{\em J. Chem. Phys.} {\bf 105}, 1601--1608 (1996).

\bibitem{los_rios}
C.~Maffi, M.~Baiesi, L.~Casetti, F.~Piazza and P.~De Los Rios,
``First-order coil-globule transition driven by vibrational entropy'',
{\em Nat. Commun.} {\bf 3}, 1065 (2012).

\bibitem{creighton}
T.~E.~Creighton, {\em Proteins: structure and molecular properties}.
\newblock New York, Ed. W.~H.~Freeman, 1993.

\bibitem{bahar_jernigan_dill}
I.~Bahar, R.~L.~Jernigan and K.~A.~Dill, {\em Protein Actions}.
\newblock Garland Science, Taylor \& Francis Group, 2017.

\bibitem{lesk}
A.~M. Lesk, {\em Introduction to Protein Science: Architecture, function and
genomics}.
\newblock Oxford University Press, 2~ed., 2004.

\bibitem{comment}
All-atoms simulations including side-chains have provided invaluable
insights into proteins. Please see the three books above for details.

\bibitem{crick}
F. Crick, {\em What mad pursuit}.
\newblock Basic Books, New York, 1988.

\end{thebibliography}

\end{document}